\documentclass[journal]{vgtc}                     % final (journal style)
\vgtccategory{Research}

\vgtcpapertype{please specify}

\title{UADAPy: An Uncertainty-Aware Visualization and Analysis Toolbox}

\author{%
\authororcid{Patrick Paetzold}{0000-0002-1315-4602},
\authororcid{David Hägele}{0000-0002-2679-6882},
\authororcid{Marina Evers}{0000-0003-3904-5065},
\authororcid{Daniel Weiskopf}{0000-0003-1174-1026},
  and \authororcid{Oliver Deussen}{0000-0001-5803-2185}
}

\authorfooter{
  \item
  	David Hägele, Marina Evers, and Daniel Weiskopf are with University of Stuttgart.
  	E-mail:\\ \{Marina.Evers, David.Haegele, Daniel.Weiskopf\}@visus.uni-stuttgart.de.
  \item
  	Patrick Paetzold, Oliver Deussen are with University of Konstanz.\\
  	E-mail: \{patrick.paetzold, oliver.deussen\}@uni-konstanz.de.
}

\abstract{%
   Current research provides methods to communicate uncertainty and adapts classical algorithms of the visualization pipeline to take the uncertainty into account. Various existing visualization frameworks include methods to present uncertain data but do not offer transformation techniques tailored to uncertain data. Therefore, we propose a software package for uncertainty-aware data analysis in Python (UADAPy)
  offering methods for uncertain data along the visualization pipeline. 
  We aim to provide a platform that is the foundation for further integration of uncertainty algorithms and visualizations. It provides common utility functionality to support research in uncertainty-aware visualization algorithms and makes state-of-the-art research results accessible to the end user.
  The project is available at \url{https://github.com/UniStuttgart-VISUS/uadapy}.
  
}


\teaser{
  \centering
  \includegraphics[width=\linewidth, alt={A view of a city with buildings peeking out of the clouds.}]{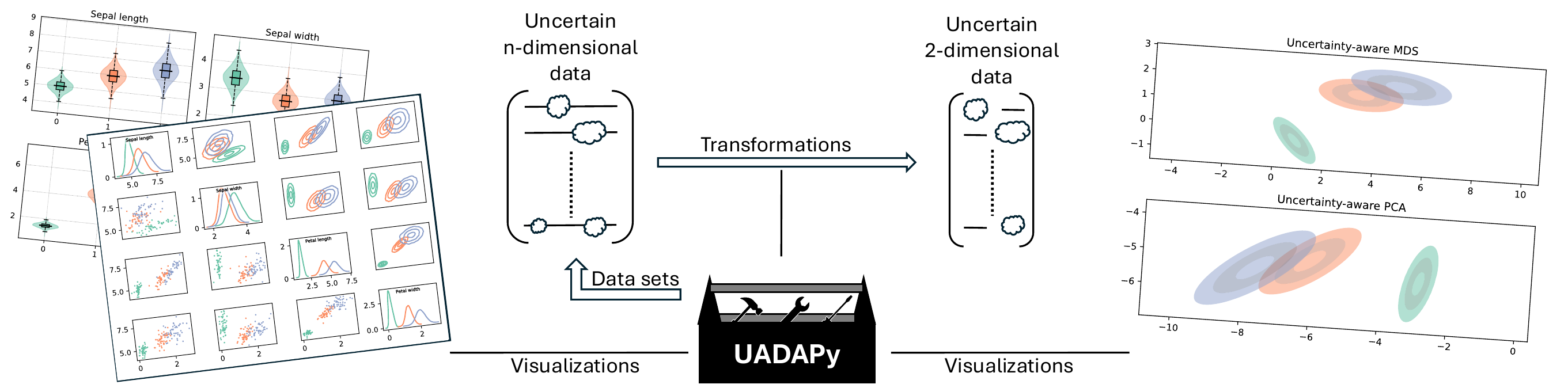}
  \caption{%
  	The UADAPy software package is a toolbox providing high-dimensional uncertain sample data sets, uncertainty-aware data transformations and analysis methods, and visualization methods tailored to show uni- and multivariate sets of probability distributions.%
  }
  \label{fig:teaser}
}

\graphicspath{{figs/}{figures/}{pictures/}{images/}{./}} % where to search for the images

\usepackage{tabu}                      % only used for the table example
\usepackage{booktabs}                  % only used for the table example
\usepackage{lipsum}                    % used to generate placeholder text
\usepackage{mwe}                       % used to generate placeholder figures
\usepackage{mathptmx}                  % use matching math font

\begin{document}

%%%%%%%%%%%%%%%%%%%%%%%%%%%%%%%%%%%%%%%%%%%%%%%%%%%%%%%%%%%%%%%%
%%%%%%%%%%%%%%%%%%%%%% START OF THE PAPER %%%%%%%%%%%%%%%%%%%%%%
%%%%%%%%%%%%%%%%%%%%%%%%%%%%%%%%%%%%%%%%%%%%%%%%%%%%%%%%%%%%%%%%

\firstsection{Introduction}

\maketitle

Adapting state-of-the-art analysis and visualization methods for uncertain data can be complicated, for example, due to the lack of ready-to-use implementations.
We want to lower the effort for users from non-visualization domains as well as visualization experts to easily adapt such novel methods for uncertain data.
Specifically, we present the software package UADAPy for analyzing and visualizing multivariate uncertain data, where uncertainties are expressed as probability distributions.
Often, uncertainty-propagating analysis methods come in combination with a custom visualization method that is capable of representing uncertainties. 
Only providing either analysis algorithms or visualization methods is not helpful, which is why we cover multiple stages of the visualization/analysis pipeline, i.e., \emph{data modeling, data transformation,} and \emph{visual mapping}, supporting the propagation of uncertainty.
The Python programming language is one of the most commonly used languages for data analysis. Therefore, we propose a Python library that provides existing uncertainty-aware algorithms and visualizations and acts as a starting point for developing new approaches.
We provide ports and reimplementations of algorithms available in other programming languages and integrate them within our simple and consistent API. Currently, we have integrated uncertainty-aware dimensionality reduction and visualization methods.
Instead of reinventing the wheel, we make use of common plotting and data analysis libraries 
to ease the integration of our package into user's analysis workflow.
Since many visualization methods, e.g. scatterplot matrices, are trivial in concept, their implementation is often lengthy and done from scratch. %, which we consider boilerplate code.
We offer ready-to-use implementations of such general visualizations tailored to uncertain data modeled via probability distributions.

Multiple visualization frameworks like D3 or VTK~\cite{vtkBook, D3} provide implementations for all stages of the visualization pipeline but do not focus on uncertainty.
Several Python and R packages exist to visualize and model uncertainty. The \emph{Uncertainty Toolbox} \cite{chung2021uncertainty} mainly focuses on regression tasks resulting from machine learning problems. 
It provides various metrics for predictive uncertainty estimates, displays the results of the metrics and uncertainties using visualizations, and provides recalibration methods. 
In contrast to this regression-focused view of uncertainty, \emph{ArviZ}~\cite{arviz_2019} mainly concentrates on visualizing uncertain data in a Bayesian setting. 
It offers many visualization techniques for probability contributions and some statistical evaluation of the probability distributions. 
The R package \emph{ggdist} \cite{ggdist} provides various visualization methods for distributions and uncertain data. 
Similar to commonly used Python visualization frameworks, it does not offer techniques for data transformation. 
Most of the presented frameworks for uncertain data and probability distributions solely focus on the last step of the visualization pipeline, namely the visualization aspect itself. 
These tools do not support uncertainty-aware data transformation techniques, which take place in the preceding steps of the visualization pipeline.  

We aim to close this gap by providing an easy-to-use toolbox that bundles uncertainty-aware methods and models along the visualization pipeline, supporting the modeling of distributions and implementation of uncertainty-aware adaptations of classical data transformation algorithms like principal component analysis (PCA) \cite{PCA} and multi-dimensional scaling (MDS) \cite{MDS}. 

\section{General Concept}
\begin{figure}
    \centering
    \includegraphics{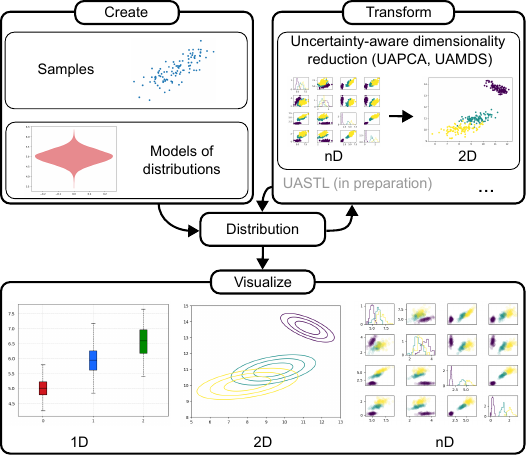}
    \caption{The distribution forms the center of our library. It can be created in different ways, transformed into other distributions by uncertainty-aware algorithms, and visualized where the choices for visual encodings depend on the dimensionality of the distribution.}
    \label{fig:concept}
\end{figure}
As a design goal for creating the library, we wanted to support the different steps of the visualization process, focusing on uncertain data, starting from loading and modeling the data over data transformation to the final visualization. However, a variety of software solutions exist for several of these steps. For the broad applicability of the unique features of our library, we closely integrate it with existing libraries such as \emph{SciPy} for data modeling or \emph{Matplotlib} for data visualization. UADAPy targets two groups of users. On the one hand, we want to support visualization researchers who develop uncertainty-aware algorithms. By providing easy access to different steps of the visualization pipeline, they do not need to reimplement the functionality of the steps unaffected by their algorithm. For example, data modeling and visualizations can be used directly when developing a new uncertainty-aware dimensionality reduction algorithm. On the other hand, we want to provide current state-of-the-art data transformation and visualization methods to end users who want to analyze uncertain data.

To achieve this goal and serve both target user groups, we follow a clear concept as shown in \cref{fig:concept}. All operations focus on the distribution class. Existing libraries, such as SciPy, provide various data structures for different distributions. However, while some functions are available to different distributions, there is no uniform interface. For flexible applicability, we provide our own distribution class that forms the center of our library. For most applications, it provides a wrapper around existing functionality but can be easily extended if other distributions are required. The distribution class provides access to key characteristics such as mean and covariance and supports drawing samples and computing probability densities independent of the underlying distribution. As shown in \cref{fig:concept}, a distribution object can be created based on data. Currently, we support the creation of distribution objects from samples through kernel density estimation or use SciPy models such as the multivariate normal distribution.

Common uncertainty-aware data transformations convert the data from one distribution to another to support uncertainty propagation. For this purpose, data transformations typically take single or multiple distribution objects as input and return a list of distribution objects, possibly with different dimensionality. This provides large flexibility in the combination of different transformations. In case transformation methods do not apply to arbitrary distributions, this is not handled on the interface level but internally by the transformation methods, allowing for specialized implementations for specific distribution types if necessary. Distributions can then be shown using different visualizations. Our visualization methods do not differentiate between the types of distribution. Instead, only the dimensionality of the distribution defines the availability of the visualization options. While the visualizations heavily build on existing libraries, our goal is to provide a high-level interface such that sets of distributions can be visualized in a single line of code. While this comes at the cost of customizability, it lowers the barrier to obtaining a first view of the data and supports rapid prototyping in the development process of new algorithms.

\section{Features}

Our proposed toolkit currently includes three main aspects. First, it includes example datasets used in various publications focusing on uncertainty visualization to provide easy entry to our library.

Second, it provides several methods to transform high-dimensional data to a low-dimensional space. For visualization purposes, it is common to apply dimensionality reduction from an $n$-dimensioanl space to only two dimensions. This is oftentimes one of the first steps of the visualization pipeline. As we aim to support $n$-dimensional probability distributions, we have already implemented two commonly used algorithms to embed high-dimensional probability distributions in 2D. UAPCA \cite{ua-pca} is an uncertainty-aware variant of the linear dimensionality reduction method PCA. Similarly, we also implemented UAMDS \cite{ua-mds}, a variant of MDS that applies to Gaussian distributions. The currently implemented techniques, such as uncertainty-aware dimensionality reduction techniques, propagate the uncertainty from the high-dimensional space to the low-dimensional embedding.

Third, as the last step of the pipeline, we support visualizing probability distributions. For bivariate distributions, we provide classical techniques like scatter plots as sample-based techniques. To provide a more aggregated representation of distributions, we support isolines and isobands as visualization techniques to show specific quantiles.
We offer box plots, violin plots, strip plots, and swarm plots to show one-dimensional summary statistics of the distributions.
If a distribution is more than two-dimensional, it is challenging to visualize.
The user can use small multiples, such as a plot matrix that shows
all pairs of dimensions as bivariate plots (e.g. scatter or contour plots), or each dimension separately as univariate plots (e.g. box or violin plots).
Alternatively, the provided dimensionality reduction methods can be applied to get a 2-dimensional probability distribution for display. 

\section{Discussion and Future Work}
Uncertainty visualization is an active research field with many algorithmic and visual advances. To make uncertainty-aware adaptations of the individual stages of the visualization pipeline more easily accessible, we propose a new analysis and visualization toolbox for uncertain data that incorporates uncertainty-aware data transformation methods as well as uncertainty visualizations. Gaussian processes are often applied to model uncertain time series. In the future, we aim to additionally support the analysis and visualization of uncertain time series.
Therefore, we strive to include the uncertainty-aware adaptation of the Seasonal-Trend Decomposition (STL)~\cite{USTL} as a further tool. To generate the visualizations, we use the commonly used plotting library \emph{Matplotlib}. As its interaction capabilities are limited, we plan to extend the visualizations to other plotting frameworks like %\emph{seaborn}, 
\emph{Bokeh} or \emph{Plotly}. Our current primary focus lies in implementing the features in Python. In the future, we plan to provide bindings to \texttt{}{R}, as it is commonly used in statistics and other disciplines and, therefore, would lead to a broader adoption of uncertainty-aware data analysis methods. We hope to establish a foundational platform where further uncertainty-aware algorithms can easily be included. Currently, our toolkit is in an early phase of development and is available at \url{https://github.com/UniStuttgart-VISUS/uadapy}. It is also provided at \url{https://pypi.org/} to make it easily installable via \texttt{pip}.

%% if specified like this the section will be omitted in review mode
\acknowledgments{This work was funded by the Deutsche Forschungsgemeinschaft (DFG,
German Research Foundation)—Project ID 251654672—TRR 161
(Project A01). In addition, the authors thank Nikhil Bhavikatti and Ruben Bauer for helping in the implementation of the library.}

\bibliographystyle{abbrv-doi-hyperref}

\bibliography{template}

\begin{thebibliography}{10}

\bibitem{D3}
M.~Bostock, V.~Ogievetsky, and J.~Heer.
\newblock D³ data-driven documents.
\newblock {\em IEEE Transactions on Visualization and Computer Graphics}, 17(12):2301--2309, 2011. \href{https://doi.org/10.1109/TVCG.2011.185}
{doi: {{%
10\hspace{.1pt}\discretionary{.}{%
}{.}\hspace{.4pt}1109\discretionary{/}{%
}{/}TVCG\hspace{.1pt}\discretionary{.}{%
}{.}\hspace{.4pt}2011\hspace{.1pt}\discretionary{.}{%
}{.}\hspace{.4pt}185}}}


\bibitem{chung2021uncertainty}
Y.~Chung, I.~Char, H.~Guo, J.~Schneider, and W.~Neiswanger.
\newblock Uncertainty toolbox: an open-source library for assessing, visualizing, and improving uncertainty quantification.
\newblock {\em arXiv preprint arXiv:2109.10254}, 2021.

\bibitem{ua-pca}
J.~Görtler, T.~Spinner, D.~Streeb, D.~Weiskopf, and O.~Deussen.
\newblock Uncertainty-aware principal component analysis.
\newblock {\em IEEE Transactions on Visualization and Computer Graphics}, 26(1):822--831, 2020. \href{https://doi.org/10.1109/TVCG.2019.2934812}
{doi: {{%
10\hspace{.1pt}\discretionary{.}{%
}{.}\hspace{.4pt}1109\discretionary{/}{%
}{/}TVCG\hspace{.1pt}\discretionary{.}{%
}{.}\hspace{.4pt}2019\hspace{.1pt}\discretionary{.}{%
}{.}\hspace{.4pt}2934812}}}


\bibitem{ua-mds}
D.~Hägele, T.~Krake, and D.~Weiskopf.
\newblock Uncertainty-aware multidimensional scaling.
\newblock {\em IEEE Transactions on Visualization and Computer Graphics}, 29(1):23--32, 2023. \href{https://doi.org/10.1109/TVCG.2022.3209420}
{doi: {{%
10\hspace{.1pt}\discretionary{.}{%
}{.}\hspace{.4pt}1109\discretionary{/}{%
}{/}TVCG\hspace{.1pt}\discretionary{.}{%
}{.}\hspace{.4pt}2022\hspace{.1pt}\discretionary{.}{%
}{.}\hspace{.4pt}3209420}}}


\bibitem{ggdist}
M.~Kay.
\newblock ggdist: Visualizations of distributions and uncertainty in the grammar of graphics.
\newblock {\em IEEE Transactions on Visualization and Computer Graphics}, 30(1):414--424, 2024. \href{https://doi.org/10.1109/TVCG.2023.3327195}
{doi: {{%
10\hspace{.1pt}\discretionary{.}{%
}{.}\hspace{.4pt}1109\discretionary{/}{%
}{/}TVCG\hspace{.1pt}\discretionary{.}{%
}{.}\hspace{.4pt}2023\hspace{.1pt}\discretionary{.}{%
}{.}\hspace{.4pt}3327195}}}


\bibitem{USTL}
T.~Krake, D.~Klötzl, D.~Hägele, and D.~Weiskopf.
\newblock Uncertainty-aware seasonal-trend decomposition based on loess.
\newblock {\em IEEE Transactions on Visualization and Computer Graphics}, pp. 1--16, 2024. \href{https://doi.org/10.1109/TVCG.2024.3364388}
{doi: {{%
10\hspace{.1pt}\discretionary{.}{%
}{.}\hspace{.4pt}1109\discretionary{/}{%
}{/}TVCG\hspace{.1pt}\discretionary{.}{%
}{.}\hspace{.4pt}2024\hspace{.1pt}\discretionary{.}{%
}{.}\hspace{.4pt}3364388}}}


\bibitem{MDS}
J.~Kruskal and M.~Wish.
\newblock Multidimensional scaling, 1978. \href{https://doi.org/10.4135/9781412985130}
{doi: {{%
10\hspace{.1pt}\discretionary{.}{%
}{.}\hspace{.4pt}4135\discretionary{/}{%
}{/}9781412985130}}}


\bibitem{arviz_2019}
R.~Kumar, C.~Carroll, A.~Hartikainen, and O.~Martin.
\newblock {ArviZ} a unified library for exploratory analysis of {B}ayesian models in python.
\newblock {\em Journal of Open Source Software}, 4(33):1143, 2019. \href{https://doi.org/10.21105/joss.01143}
{doi: {{%
10\hspace{.1pt}\discretionary{.}{%
}{.}\hspace{.4pt}21105\discretionary{/}{%
}{/}joss\hspace{.1pt}\discretionary{.}{%
}{.}\hspace{.4pt}01143}}}


\bibitem{PCA}
K.~Pearson.
\newblock {LIII}. {O}n lines and planes of closest fit to systems of points in space.
\newblock {\em The London, Edinburgh, and Dublin Philosophical Magazine and Journal of Science}, 2(11):559--572, 1901. \href{https://doi.org/10.1080/14786440109462720}
{doi: {{%
10\hspace{.1pt}\discretionary{.}{%
}{.}\hspace{.4pt}1080\discretionary{/}{%
}{/}14786440109462720}}}


\bibitem{vtkBook}
W.~Schroeder, K.~Martin, and B.~Lorensen.
\newblock {\em The Visualization Toolkit (4th ed.)}.
\newblock Kitware, 2006.

\end{thebibliography}

\appendix % You can use the `hideappendix` class option to skip everything after \appendix

\end{document}